\documentclass[12pt]{article}
\usepackage{epsfig}

\usepackage{mathrsfs}
\usepackage[T1]{fontenc}
\usepackage{mathpazo}
\usepackage{setspace}
\usepackage{amsfonts}
\usepackage{amssymb}
\usepackage{amsmath}
\usepackage{epsfig}
\usepackage{latexsym}
\usepackage{color}
\usepackage{graphicx}
\usepackage{nicefrac}
\usepackage[latin1]{inputenc}
\usepackage{pstricks}
\usepackage{slashed}
\usepackage{multirow}

\def\hybrid{\topmargin -20pt    \oddsidemargin 0pt
        \headheight 0pt \headsep 0pt
        \textwidth 6.25in       
        \textheight 9.5in       
        \marginparwidth .875in
        \parskip 5pt plus 1pt   \jot = 1.5ex}

\hybrid

\def\baselinestretch{1.2}

\catcode`\@=11

\def\marginnote#1{}
\newcount\hour
\newcount\minute
\newtoks\amorpm
\hour=\time\divide\hour by60 \minute=\time{\multiply\hour by60
\global\advance\minute by-\hour}
\edef\standardtime{{\ifnum\hour<12 \global\amorpm={am}%
        \else\global\amorpm={pm}\advance\hour by-12 \fi
        \ifnum\hour=0 \hour=12 \fi
        \number\hour:\ifnum\minute<10 0\fi\number\minute\the\amorpm}}
\edef\militarytime{\number\hour:\ifnum\minute<10
0\fi\number\minute}

\def\draftlabel#1{{\@bsphack\if@filesw {\let\thepage\relax
   \xdef\@gtempa{\write\@auxout{\string
      \newlabel{#1}{{\@currentlabel}{\thepage}}}}}\@gtempa
   \if@nobreak \ifvmode\nobreak\fi\fi\fi\@esphack}
        \gdef\@eqnlabel{#1}}
\def\@eqnlabel{}
\def\@vacuum{}
\def\draftmarginnote#1{\marginpar{\raggedright\scriptsize\tt#1}}

\def\draft{\oddsidemargin -.5truein
        \def\@oddfoot{\sl preliminary draft \hfil
        \rm\thepage\hfil\sl\today\quad\militarytime}
        \let\@evenfoot\@oddfoot \overfullrule 3pt
        \let\label=\draftlabel
        \let\marginnote=\draftmarginnote
   \def\@eqnnum{(\theequation)\rlap{\kern\marginparsep\tt\@eqnlabel}%
\global\let\@eqnlabel\@vacuum}  }

\def\preprint{\twocolumn\sloppy\flushbottom\parindent 2em
        \leftmargini 2em\leftmarginv .5em\leftmarginvi .5em
        \oddsidemargin -.5in    \evensidemargin -.5in
        \columnsep .4in \footheight 0pt
        \textwidth 10.in        \topmargin  -.4in
        \headheight 12pt \topskip .4in
        \textheight 6.9in \footskip 0pt
        \def\@oddhead{\thepage\hfil\addtocounter{page}{1}\thepage}
        \let\@evenhead\@oddhead \def\@oddfoot{} \def\@evenfoot{} }

\def\numberbysection{\@addtoreset{equation}{section}
        \def\theequation{\thesection.\arabic{equation}}}

\def\underline#1{\relax\ifmmode\@@underline#1\else
        $\@@underline{\hbox{#1}}$\relax\fi}

\def\titlepage{\@restonecolfalse\if@twocolumn\@restonecoltrue\onecolumn
     \else \newpage \fi \thispagestyle{empty}\c@page\z@
        \def\thefootnote{\fnsymbol{footnote}} }

\def\endtitlepage{\if@restonecol\twocolumn \else \newpage \fi
        \def\thefootnote{\arabic{footnote}}
        \setcounter{footnote}{0}}  

\catcode`@=12 \relax

\def\figcap{\section*{Figure Captions\markboth
        {FIGURECAPTIONS}{FIGURECAPTIONS}}\list
        {Figure \arabic{enumi}:\hfill}{\settowidth\labelwidth{Figure
999:}
        \leftmargin\labelwidth
        \advance\leftmargin\labelsep\usecounter{enumi}}}
 \relax
\def\tablecap{\section*{Table Captions\markboth
        {TABLECAPTIONS}{TABLECAPTIONS}}\list
        {Table \arabic{enumi}:\hfill}{\settowidth\labelwidth{Table
999:}
        \leftmargin\labelwidth
        \advance\leftmargin\labelsep\usecounter{enumi}}}
 \relax
\def\reflist{\section*{References\markboth
        {REFLIST}{REFLIST}}\list
        {[\arabic{enumi}]\hfill}{\settowidth\labelwidth{[999]}
        \leftmargin\labelwidth
        \advance\leftmargin\labelsep\usecounter{enumi}}}
 \relax
\makeatletter
\newcounter{pubctr}
\def\publist{\@ifnextchar[{\@publist}{\@@publist}}
\def\@publist[#1]{\list
        {[\arabic{pubctr}]\hfill}{\settowidth\labelwidth{[999]}
        \leftmargin\labelwidth
        \advance\leftmargin\labelsep
        \@nmbrlisttrue\def\@listctr{pubctr}
        \setcounter{pubctr}{#1}\addtocounter{pubctr}{-1}}}
\def\@@publist{\list
        {[\arabic{pubctr}]\hfill}{\settowidth\labelwidth{[999]}
        \leftmargin\labelwidth
        \advance\leftmargin\labelsep
        \@nmbrlisttrue\def\@listctr{pubctr}}}
 \relax
\makeatother
\newskip\humongous \humongous=0pt plus 1000pt minus 1000pt

\newif\ifdtup

\relax

\def\be{\begin{equation}}
\def\ee{\end{equation}}
\def\ba{\begin{eqnarray}}
\def\ea{\end{eqnarray}}

\def\no{\noindent}

\def\IR{\relax{\rm I\kern-.18em R}}

\def\IR{\relax{\rm I\kern-.18em R}}
\def\inv{^{\raise.15ex\hbox{${\scriptscriptstyle -}$}\kern-.05em 1}}

\begin{document}

\renewcommand{\theequation}{\thesection.\arabic{equation}}

\newcommand{\beq}{\begin{equation}}
\newcommand{\eeq}[1]{\label{#1}\end{equation}}
\newcommand{\ber}{\begin{eqnarray}}
\newcommand{\eer}[1]{\label{#1}\end{eqnarray}}
\newcommand{\eqn}[1]{(\ref{#1})}
\begin{titlepage}
\begin{center}

\vskip -.1 cm
\hfill September 2010\\

\vskip .6in

{\large \bf Gradient flows and instantons at a Lifshitz point}\footnote{Based
on lectures delivered at the following workshops and conferences: {\em Membranes,
Minimal Surfaces and Matrix Limits}, 19--21 October 2009, Golm, Germany;
{\em 7th Spring School and Workshop on Quantum Field Theory and Hamiltonian
Systems}, 10--15 May 2010, Calimanesti, Romania; {\em Recent Developments in
Gravity: NEB14}, 8--11 June 2010, Ioannina, Greece; {\em Dynamics in Samos},
31 August -- 3 September 2010, Karlovassi, Greece. Contribution to
appear in Journal of Physics Conference Series.}

\vskip 0.6in

{\bf Ioannis Bakas} \vskip 0.2in
{\em Department of Physics, University of Patras \\
GR-26500 Patras, Greece\\
\vskip 0.2in
\footnotesize{\tt bakas@ajax.physics.upatras.gr}}\\

\end{center}

\vskip 0.8in

\centerline{\bf Abstract} \no
I provide a broad framework to embed gradient flow equations in non--relativistic
field theory models that exhibit anisotropic scaling. The prime example
is the heat equation arising from a Lifshitz scalar field theory; other examples
include the Allen--Cahn equation that models the evolution of phase boundaries.
Then, I review recent results reported in arXiv:1002.0062 describing instantons
of Ho\v{r}ava--Lifshitz gravity as eternal solutions of certain geometric flow
equations on 3--manifolds. These instanton solutions are in general chiral
when the anisotropic scaling exponent is $z=3$. Some general connections with the
Onsager--Machlup theory of non--equilibrium processes are also briefly discussed
in this context. Thus, theories of Lifshitz type in $d+1$ dimensions can be used
as off--shell toy models for dynamical vacuum selection of relativistic field
theories in $d$ dimensions.

\vfill
\end{titlepage}
\eject

\def\baselinestretch{1.2}
\baselineskip 16 pt \noindent
\section{Preliminaries: Lifshitz theories}
\setcounter{equation}{0}

Let us consider a point particle system with configuration space $\mathbb{Q}$
and a system of local coordinates $q_I$ that describe its physical degrees of freedom.
We further assume that $\mathbb{Q}$ is endowed with a metric ${\cal O}^{IJ}$ which is
symmetric and non--degenerate so that the inverse metric ${\cal O}_{IJ}$ defined
through ${\cal O}^{IK} {\cal O}_{KJ} = \delta_J^I$ exists everywhere. The action
comprises of two terms taking the difference of the kinetic and potential energy
\begin{equation}
S = {1 \over 2} \int dt \sum_{I, J} \left({dq_I \over dt} {\cal O}^{IJ} {dq_J \over dt}
- {\partial W \over \partial q_I} {\cal O}_{IJ} {\partial W \over \partial q_J}
\right) ,
\label{action}
\end{equation}
where we also assume that the potential is derived from a superpotential $W$ that is
a function on $\mathbb{Q}$. This particular choice of potential is synonymous to
having detailed balance condition in the system. Here, we will only consider systems
with positive definite metric ${\cal O}^{IJ}$ so that the potential term is manifestly
positive definite and its minima coincide with the critical points of $W$, i.e.,
$\partial W / \partial q_I = 0$, $\forall ~ I$. Then, the
ground states of the system describe a particle sitting still at the minima of the
potential and there can be more than one degenerate vacua depending upon the choice
of $W$. It is often interesting to consider systems with configuration space having
indefinite metric ${\cal O}^{IJ}$, although some of the nice properties that are
discussed in the following will be generally missing.

When the dimension of the configuration space $\mathbb{Q}$ is finite, we have
an ordinary point particle system, which is not particularly interesting by itself
but it is a useful guide for some of the constructions described below.
Here, we focus entirely on infinite dimensional systems determined by the action
\eqn{action} and consider only those cases that
$\mathbb{Q}$ is the configuration space of a relativistic field theory in $d$
Euclidean space--time dimensions. The superpotential $W$ is a functional over the
field space, which for all practical purposes is taken to be the classical
action of the relativistic field theory in $d$ dimensions. This construction may
look odd at first sight, as the corresponding action \eqn{action} describes a
non--relativistic field theory in $d+1$ space--time dimensions with $t$ being the
physical time coordinate, but, in fact, it provides our main framework. The
resulting field theories are called models of Lifshitz type, \cite{lifsh}, and
they exhibit anisotropic scaling (typically it is $z=2$ for theories with $W$
that contains up to two derivative terms, but there can be more general examples
as will be seen later).

The prime example of a Lifshitz theory in $d+1$ space--time dimensions is
provided by taking $\mathbb{Q}$ to be the configuration space of a relativistic
free field theory in $d$ Euclidean dimensions. For this it is appropriate to set
$q_I (t) = \varphi (t, x)$, $q_J (t) = \varphi (t, x^{\prime})$, ${\cal O}^{IJ} =
\delta (x-x^{\prime}) = {\cal O}_{IJ}$ and choose
\begin{equation}
W[\varphi] = {1 \over 2} \int d^d x \left(\nabla \varphi \right)^2
\end{equation}
so that $\partial W / \partial q_I = \delta W / \delta \varphi(x) = - \nabla^2
\varphi (x)$. Then, by integration over $x^{\prime}$ (equivalently, summation over
$J$), the point particle action \eqn{action} takes the form
\begin{equation}
S = {1 \over 2} \int dt ~ d^d x \Big[\left(\partial_{\rm t} \varphi \right)^2 -
\left(\nabla^2 \varphi \right)^2 \Big]
\label{lif1}
\end{equation}
and describes a non-relativistic field theory with anisotropic scaling $z=2$
known as Lifshitz scalar theory. The resulting equations of motion exhibit scale
invariance under the transformation $x \rightarrow a x$, $t \rightarrow a^z t$ with
exponent $z=2$; it should be contrasted to the relativistic scalar free field theory
in $d+1$ space-time dimensions whose Lagrangian density is $(\partial_{\rm t} \varphi)^2
- (\nabla \varphi)^2$ and the corresponding scaling exponent is $z=1$.

A simple variant of the model is obtained by considering the action of a
self--interacting relativistic scalar field in $d$ Euclidean dimensions as
superpotential functional,
\begin{equation}
W[\varphi] = \int d^d x \Big[{1 \over 2} \left(\nabla \varphi \right)^2 + V(\varphi) \Big] ~,
\label{soura1}
\end{equation}
in which case the corresponding Lifshitz theory in $d+1$ space--time dimensions takes
the form
\begin{equation}
S = {1 \over 2} \int dt ~ d^d x \Big[\left(\partial_{\rm t} \varphi \right)^2 -
\left(\nabla^2 \varphi - V^{\prime} (\varphi) \right)^2 \Big] ~,
\label{lif2}
\end{equation}
where $V^{\prime} (\varphi) = \partial V / \partial \varphi$. This particular field theory
will be discussed later in more detail as it connects with some interesting
mathematical problems of current interest. One may also consider more
complicated examples using multi--component scalar field models.

Another characteristic class of Lifshitz models is provided by vector field theories
with anisotropic scaling in space and time. As before, it is appropriate to use the
action \eqn{action} as starting point and identify $q_I$ with gauge fields $A_i (x)$
that live in $d$--dimensional Euclidean space, letting $i = 1, 2, \cdots , d$.
The metric ${\cal O}^{IJ}$ is provided by the standard Riemannian metric
${\rm Tr}(T_a T_b) \delta (x-x^{\prime})$ in the space of all $d$--dimensional gauge
field configurations $A_i = A_i^a T_a$ with values in a Lie algebra
$[T_a , ~ T_b] = i{f_{ab}}^c T_c$. Then, it is natural to define a Lifshitz gauge
field theory in terms of the action, \cite{horava1},
\begin{equation}
S = {1 \over 2} \int dt ~ d^dx \Big[{\rm Tr} (E_i E^i) - {1 \over g^2} {\rm Tr}
\left((\partial_i F^{ik}) (\partial_j {F^j}_k) \right) \Big] ~,
\label{lif3}
\end{equation}
where
\begin{equation}
E_i = \partial_{\rm t} A_i ~, ~~~~~~ F_{ij} = \partial_i A_j - \partial_j A_i -
i[A_i , ~ A_j]
\end{equation}
provide the electric and magnetic fields in $d+1$ space--time dimensions, respectively,
in the axial gauge $A_0 (t, x) =0$. This action follows from \eqn{action}
using the superpotential functional
\begin{equation}
W = {1 \over 4g^2} \int d^d x ~ {\rm Tr} (F_{ij} F^{ij})
\label{soura2}
\end{equation}
associated to relativistic Yang--Mills theory in $d$ Euclidean dimensions, and,
therefore, the resulting Lifshitz vector theory exhibits anisotropic scaling with
exponent $z=2$.

Finally, we consider Lifshitz theories of geometric type associated to tensors of
rank 2, which were introduced in the literature as alternative theories of gravity
and they became known as Ho\v{r}ava--Lifshitz gravities,
\cite{horava2, horava}. They are defined in arbitrary dimensions
by assuming that space--time is of the form $M_{d+1} = \mathbb{R} \times \Sigma_d$ and
the metric admits the ADM (Arnowitt--Deser--Misner) decomposition, \cite{adm}
\begin{equation}
\mathrm{d}s^2 = -N^2 \mathrm{d}t^2 +g_{ij}\left(\mathrm{d}x^i +N^i \mathrm{d}t\right)
\left(\mathrm{d}x^j +N^j \mathrm{d}t\right) .
\end{equation}
The metric on the spatial slices $\Sigma_d$ is $g_{ij}$, whereas $N$ and $N^i$ are the
lapse and shift functions, respectively, which depend on all space--time coordinates,
in general. The infinite dimensional space of all Riemannian metrics $g_{ij}$ is called
superspace and it is endowed with a metric
\begin{equation}
\label{dwm}
\mathcal{G}^{ijk\ell}=\frac{1}{2}\left(g^{ik}g^{j\ell}+g^{i\ell}g^{jk}\right)-\lambda
g^{ij}g^{k\ell}
\end{equation}
that generalizes the standard DeWitt metric, \cite{witt}, using an arbitrary parameter
$\lambda$ (other than $1$). Here, for simplicity, we are suppressing the delta functions
that are needed to integrate over space. The inverse metric is
\begin{equation}
\mathcal{G}_{ijk\ell}=\frac{1}{2}\left(g_{ik}g_{j\ell}+g_{i\ell}g_{jk}\right)-
\frac{\lambda}{d\lambda-1} g_{ij}g_{k\ell}
\end{equation}
so that
\begin{equation}
\mathcal{G}^{ijk\ell}\mathcal{G}_{k\ell mn}= \frac{1}{2}(\delta^i_m \delta^j_n +
\delta^i_n \delta^j_m ) ~.
\end{equation}
The metric in superspace is positive definite provided that $\lambda < 1/d$, but otherwise
it is arbitrary.

The action of Ho\v{r}ava--Lifshitz gravity in $d+1$ dimensions is written as a
sum of kinetic and potential terms. By identifying $q_I$ with $g_{ij}$ and ${\cal O}^{IJ}$
with $\mathcal{G}^{ijk\ell}$, the action \eqn{action} satisfying detailed balance takes
the form, \cite{horava2, horava},
\begin{equation}
S  =  {2 \over \kappa^2} \int dt d^dx \sqrt{g} ~ N ~ K_{ij} \mathcal{G}^{ijk\ell}
K_{k\ell} - {\kappa^2 \over 2} \int dt d^dx \sqrt{g} ~ N ~ E^{ij} \mathcal{G}_{ijk\ell}
E^{k\ell} ~,
\label{lif4}
\end{equation}
where $K_{ij}$ is the second fundamental form measuring the extrinsic curvature
of the spatial slices $\Sigma_d$ at constant $t$ (playing the role of momentum conjugate
to the metric $g_{ij}$),
\begin{equation}
K_{ij}=\frac{1}{2N}\left(\partial _t g_{ij}-\nabla_iN_j-\nabla_jN_i\right)
\end{equation}
and
\begin{equation}
\label{eom}
E^{ij}=-{1 \over 2 \sqrt{g}}\frac{\delta W[g]}{\delta g_{ij}} ~.
\end{equation}
The gravitational coupling of the theory in $d+1$ dimensions is $\kappa$.

Note that the kinetic term contains two time derivatives of the metric $g_{ij}$, and,
as such, it is identical to ordinary general relativity in canonical form (though
$\lambda$ is taken arbitrary here). The potential term is different, however, as it is
derived from a superpotential functional $W [g]$, which is typically chosen to be the
Euclidean action of a relativistic gravitational theory in $d$ dimensions. If $W$ is
the Einstein--Hilbert action, the resulting Ho\v{r}ava--Lifshitz theory will have
anisotropic scaling $z=2$, but if $W$ is the action of a covariant higher derivative
gravitational theory in $d$ dimensions the scaling exponent will be $z>2$.
Examples of this kind will be discussed later. Finally, note that the action \eqn{lif4}
is not generally covariant, since by construction it is only invariant under the
restricted set of foliation preserving
diffeomorphisms of the space--time $\mathbb{R} \times \Sigma_d$. In the following, we
restrict attention to the so called projectable version of Ho\v{r}ava-Lifshitz
theory, meaning that the lapse function $N$ associated to the freedom of time
reparametrization is restricted to be a function of $t$, whereas the shift functions
$N_i$ associated to diffeomorphisms of $\Sigma_d$ can depend on all space-time
coordinates. Also, in view of the applications that will be discussed next, we choose
\begin{equation}
N(t) = 1 ~, ~~~~~~ N_i (t, x) = 0
\label{spelcho}
\end{equation}
without great loss of generality. We will indicate later how the lapse and shift
functions can be reinstated into the equations of motion.

\section{Gradient flows as Euclidean solutions}
\setcounter{equation}{0}

Let us now consider the Euclidean form of the action \eqn{action} obtained by
Wick rotation $t \rightarrow it$ and construct solutions that are applicable to
all Lifshitz type theories. We have
\begin{equation}
S_{\rm Eucl.} = {1 \over 2} \int dt \sum_{I, J} \left({dq_I \over dt} {\cal O}^{IJ}
{dq_J \over dt} + {\partial W \over \partial q_I} {\cal O}_{IJ}
{\partial W \over \partial q_J} \right) ,
\end{equation}
which can be alternatively written as follows
\begin{equation}
S_{\rm Eucl.} = {1 \over 2} \int dt \left({dq_I \over dt} \mp {\cal O}_{IK}
{\partial W \over \partial q_K} \right) {\cal O}^{IJ}
\left({dq_J \over dt} \mp {\cal O}_{JL}
{\partial W \over \partial q_L} \right) \pm \int dt ~ {dq_I \over dt}
{\partial W \over \partial q_I}
\label{action2}
\end{equation}
by completing the square. Here, summation is implicity assumed over repeated
indices in order to simplify the presentation. For now and later use we
also consider the action
\begin{equation}
S_{\rm Eucl.}^{\prime} = {1 \over 2} \int dt \left({dq_I \over dt} \mp
{\cal O}_{IK} {\partial W \over \partial q_K} \right) {\cal O}^{IJ}
\left({dq_J \over dt} \mp {\cal O}_{JL}
{\partial W \over \partial q_L} \right) ,
\end{equation}
which differs from \eqn{action2} by boundary terms as
\begin{equation}
S_{\rm Eucl.} = S_{\rm Eucl.}^{\prime} \pm \int dt ~ {dW \over dt} ~.
\label{differe}
\end{equation}

Since we are only considering positive definite metrics ${\cal O}^{IJ}$, the action
$S_{\rm Eucl.}^{\prime}$ is bounded from below by zero. Thus, minima of the
action are provided by special configurations that satisfy the system of
first order equations
\begin{equation}
{d q_I \over dt} = \pm {\cal O}_{IJ} {\partial W \over \partial q_I} ~.
\label{gradie}
\end{equation}
These are gradient flow equations for the variables $q_I (t)$ derived from the
superpotential $W[q]$. They also yield solutions of the equations of motion
following from the action
$S_{\rm Eucl.}$, since the difference is a total derivative term \eqn{differe}
that certainly cannot affect the classical equations of motion. However, this
boundary term is needed to make the variational problem well--posed and will be
treated more carefully in the next section for, otherwise, the action
$S_{\rm Eucl.}$ may become infinite for solutions that become singular
at a finite instant of time.
In any case, the fixed points of the flow equations \eqn{gradie} represent
static solutions, where the point particle is sitting still at the bottom of the
potential. On the other hand, time dependent solutions of the flow equations
are more interesting to consider in general, but they are complicated; we will
present some examples that are worth studying in detail. It should also be noted
that the superpotential $W$ changes monotonically
along such flow lines when ${\cal O}^{IJ}$ is positive definite, since
\begin{equation}
{dW \over dt} = {d q_I \over dt} {\partial W \over \partial q_I} = \pm
{\partial W \over \partial q_I} {\cal O}_{IJ} {\partial W \over \partial q_J} ~,
\end{equation}
and, therefore, it provides an entropy functional for the corresponding evolution.

The simplest (but very instructive) example in the class of Lifshitz field theories
is provided by the scalar field model \eqn{lif1} in the Euclidean domain. The
corresponding gradient flow is linear and coincides with the forward or backward heat
equation
\begin{equation}
{\partial \varphi \over \partial t} = \pm \nabla^2 \varphi ~,
\end{equation}
depending on the choice of sign. Then, solutions of the heat equation in $d$ dimensions
are solutions of the Lifshitz scalar free field theory in $d+1$ Euclidean dimensions.
Note, however, that the fundamental solution of the heat equation, which
is a Gaussian function with time dependent height $1/t^{d/2}$ and width $t^{1/2}$, cannot
exist for ever since $\varphi (t, x)$ becomes singular at some finite instant of time, say
$t=0$.  This simple example illustrates the difference between finite action solutions
following from $S_{\rm Eucl.}$ and $S_{\rm Eucl.}^{\prime}$ where the boundary terms
\eqn{differe} can in fact play important role. If we were only concerned with solutions
of the flow equations that existed for a short time, the difference would have
been irrelevant. The singularities are also preventing to uplift these solutions
to Lifshitz theories that exist for all time.

These remarks are not only tailored for
the Lifshitz scalar free field theory, but, in fact, they apply to all
Lifshitz models. They also motivate the construction of instanton solutions of Lifshitz
theories that will be discussed in the next section.

The next more complicated example is provided by a self--interacting Lifshitz scalar
field as described by \eqn{lif2}. According to the general framework, Euclidean solutions
are provided by the non--linear flow equation
\begin{equation}
{\partial \varphi \over \partial t} = \pm \left(\nabla^2 \varphi - V^{\prime} (\varphi)
\right) ,
\end{equation}
which is called Allen--Cahn equation, \cite{cahn}. This equation arose first in the
literature as phenomenological model for the motion of phase boundaries by surface
tension, but it was subsequently studied in mathematics very extensively.
In this context, which is deeply connected to motion by mean curvature flow and the
problem of minimal surfaces, \cite{ilman, pino}, it seems more appropriate to
consider the closely related equation
\begin{equation}
{\partial \varphi^{\epsilon} \over \partial t} = \pm \left(\nabla^2 \varphi^{\epsilon} -
{1 \over \epsilon^2} ~ V^{\prime} (\varphi^{\epsilon})
\right) ,
\end{equation}
which follows from above by implementing the anisotropic scaling transformation of
the Lifshitz theory, $x \rightarrow x/ \epsilon $ and $t \rightarrow t/ \epsilon^2$,
and investigate this in the limit of small $\epsilon$ when the potential has two equal
wells, choosing, for example, $V(\varphi) = (\varphi^2 - \varphi_0^2)^2$. Solving the
Allen--Cahn equation is not an easy task and it provides an active area of research
in mathematics (see, for instance, \cite{aliko}, for some basic facts).

Euclidean solutions of gauge theories of Lifshitz type \eqn{lif3} are also interesting
to consider as the corresponding gradient flow assumes the form
\begin{equation}
E_i \equiv {\partial A_i \over \partial t} = \pm {1 \over g} ~ \nabla_j {F^j}_i
\end{equation}
and coincides with the so called Yang--Mills flow in $d$ dimensions. This is another very
interesting system of equations that has been studied for some time in mathematics
(e.g., \cite{struw, schl}), but many of its aspects remain open problems to this day.
Of course, one may also
consider further generalizations by combining the action functionals \eqn{soura1} and
\eqn{soura2} into a $W$ that describes the action of a relativistic Yang--Mills--Higgs
theory in $d$ Euclidean dimensions. Then, the gradient flow equations of the
corresponding Lifshitz theory will be mixture of the Allen--Cahn equation and the
Yang--Mills flow whose general properties remain to be studied (see, however, \cite{hassel}).

Finally, we come to the Euclidean solutions of geometric Lifshitz theories, such as
Ho\v{r}ava--Lifshitz gravity described by the action \eqn{lif4}. In this case, the
gradient flow equation becomes, \cite{horava2, horava}, \cite{bakas},
\begin{equation}
K_{ij} \equiv {1 \over 2} ~ {\partial g_{ij} \over \partial t} = \pm
{\kappa^2 \over 2} ~ \mathcal{G}_{ijk\ell} E^{k\ell}
\label{rouli}
\end{equation}
and gives rise to geometric flows for the metric $g_{ij}$ on $\Sigma_d$ that
depend on the choice of $W$. They can have second or higher order derivatives in
space coordinates.

More precisely, we have the following evolution equation in the general case of
Ho\v{r}ava--Lifshitz models
\begin{equation}
{\partial g_{ij} \over \partial t} = \mp
{\kappa^2 \over 2 \sqrt{g}} ~ \mathcal{G}_{ijk\ell} {\delta W \over \delta g_{k\ell}} ~,
\end{equation}
which for $W$ given by the Einstein--Hilbert action in $d > 2$ dimensions,
\begin{equation}
W [g] = {2 \over \kappa_{\rm w}^2} \int d^d x \sqrt{g} ~ R ~,
\label{hilba}
\end{equation}
it specializes to a variant of the celebrated Ricci flow equation (see, for instance,
\cite{yau})
\begin{equation}
{\partial g_{ij} \over \partial t} = \mp {\kappa^2 \over \kappa_{\rm w}^2} \left(R_{ij}
- {2 \lambda -1 \over 2(d \lambda -1)} ~ R g_{ij} \right) .
\label{riccia}
\end{equation}
It runs forward or backward in time depending on the choice of the overall sign and the
fixed points are Ricci flat metrics, $R_{ij} = 0$, independent of $\lambda$.

More general choices of $W[g]$ can also be made, depending on $d$, in which case the
exponent $z$ of anisotropic scaling can become bigger than $2$. Although the corresponding
flow equations are mathematically much more complicated for $z>2$, as they involve higher
derivative terms, they are physically better behaved for various reasons. The
simplest choice \eqn{hilba} that leads to the Ricci flow shares many similarities with the
heat equation in that the geometry on $\Sigma_d$ becomes singular at some finite
instant of time, \cite{yau}. This is actually a generic phenomenon that can render the
action $S_{\rm Eucl.}$ (but not $S_{\rm Eucl.}^{\prime}$) of $z=2$ Ho\v{r}ava--Lifshitz
gravity infinite along those flow lines ($W$ becomes infinite at the ultra--violet fixed
point of type I ancient solutions and vanishes at the singularity in $d>2$ dimensions). More
importantly, the corresponding space--time metric on $\mathbb{R} \times \Sigma_d$, which
is supposed to exist for all time, will be incomplete by the presence of singularities,
thus making such solutions totally unacceptable. In the next section
we will provide simple criteria that circumvent this problem in Lifshitz type theories
and obtain non--singular solutions, where it is appropriate.

Concluding this section, we mention that the geometric flow equations can be modified
by allowing arbitrary reparametrizations on $\Sigma_d$ generated by a vector field
$\xi_i (t, x)$ that may also depend on time. Then, the evolution \eqn{rouli} generalizes
to
\begin{equation}
{\partial g_{ij} \over \partial t} = \mp
{\kappa^2 \over 2 \sqrt{g}} ~ \mathcal{G}_{ijk\ell} {\delta W \over \delta g_{k\ell}}
+ \nabla_i \xi_j + \nabla_j \xi_i ~.
\end{equation}
The projectable version Ho\v{r}ava--Lifshitz theory with lapse and shift functions
$N(t)$ and $N_i (t, x)$ accommodates nicely this modification by choosing $\xi_i = N_i/N$.
Of course, $N(t)$ can always be set equal to $1$ by appropriate redefinition of time,
but $N_i (t, x)$ is left arbitrary, in general, since the theory is invariant under
foliation preserving diffeomorphisms of the space--time manifold
$\mathbb{R} \times \Sigma_d$. The choice \eqn{spelcho} with $N_i = 0$ amounts to taking
$\xi_i = 0$ in the flow equations.

\section{Instantons at a Lifshitz point}
\setcounter{equation}{0}

The definition of instantons in Lifshitz theories requires careful treatment of the
boundary terms arising in the Euclidean action. Returning back to it we have
\begin{equation}
S_{\rm Eucl.} = {1 \over 2} \int dt \left({dq_I \over dt} \mp
{\cal O}_{IK} {\partial W \over \partial q_K} \right) {\cal O}^{IJ}
\left({dq_J \over dt} \mp {\cal O}_{JL}
{\partial W \over \partial q_L} \right) \pm \int dt ~ {dW \over dt} ~,
\end{equation}
which for theories with positive definite metric ${\cal O}^{IJ}$ yields immediately
\begin{equation}
S_{\rm Eucl.} \geq \pm \int dt ~ {dW \over dt} ~.
\label{mitsoul}
\end{equation}
The lower bound is saturated for special configurations satisfying the gradient
flow equation
\begin{equation}
{d q_I \over dt} = \pm {\cal O}_{IJ} {\partial W \over \partial q_I} ~,
\end{equation}
which thus provide extrema of the Euclidean action and hence solutions to the
classical equations of motion as for $S_{\rm Eucl.}^{\prime}$ that was considered
before.

Note, however, that the lower bound of the action \eqn{mitsoul} need not be finite
on general grounds. All solutions of the gradient flow equations that are taken over a
finite time interval $(t_1 , ~ t_2)$ will have finite action,
\begin{equation}
S_{\rm Eucl.} = | W(t) |_{t_1}^{t_2} ~,
\end{equation}
but this does not necessarily extend smoothly to the entire line $-\infty < t < +\infty$.
What we really need is to have eternal solutions of the flow equations, which exist for
all time, and interpolate smoothly between different critical points of the
superpotential $W$. Then, the bound will have topological meaning and
the corresponding trajectories will be bona fide instantons with finite action
\begin{equation}
S_{\rm instanton} = | \Delta W | \equiv | W(t = + \infty) - W(t = - \infty) | ~.
\end{equation}
This requirement defines the notion of instantons in Lifshitz theories, \cite{bakas},
and places attention to those models that have degenerate vacua. Also, to
avoid unnecessary complications that may arise when considering Lifshitz field
theories on $\mathbb{R} \times \Sigma_d$, we implicitly assume that $\Sigma_d$
is compact without boundaries (e.g., $S^d$) so that no additional space--boundary
terms will come into play when putting lower bounds on the Euclidean action.

Let us elaborate more on the construction. First note that the lower bound of
$S_{\rm Eucl.}$ cannot be zero, since there are no periodic trajectories in Euclidean
time; if that were the case, it would have been in contradiction with the monotonicity
of $W(t)$ along the gradient flow lines. Thus, the lower bound is either positive
and finite or infinite. Second, it is not at all guaranteed that the gradient flow
equations will admit solutions that exist for sufficiently long time, as this is not
easy to prove mathematically in general; if not, the Euclidean solutions we
are considering will not be viable choices. Third, even if one can prove short time
existence of the solutions, it is not clear whether these will extend to
solutions for all time. Typically, the flow lines terminate at some finite
instant of time by developing singularities and they cannot be continued further.
Investigating the formation of singularities and their properties is a formidable
task in general and special mathematical tools need to be developed in each case
separately depending upon $W$. However, it is reasonable to expect that such
singularities, if present, will be irremovable and that $W$ can become infinite
along the corresponding flow lines; the
simple examples of the well studied heat equation and Ricci flow fully support
this point. Thus, the only systematic way to avoid configurations with infinite
Euclidean action is to consider eternal solutions of the flow equations. These are
also natural from a physical point of view, since there is no a priori reason
to have Lifshitz theories that make good sense only for finite time intervals.

In conclusion, we are only considering eternal solutions of the gradient flow
equations as viable first order Euclidean time solutions of Lifshitz theories.
Thus, to implement this program we do not need to prove the short time existence
of the flow lines, in general, nor to investigate the possible formation of
singularities in finite time. These may be interesting mathematical problems
in their own right, but we can live without them when considering instantons.
The only mathematical problems that have to be addressed in the present context
is the existence, the construction and the classification of eternal
solutions of gradient flows which are derived from a $W$ with at least two critical
points. There are many examples of Lifshitz theories that one can study in
this context, but we will confine our attention to non--relativistic gravitational
models, following \cite{bakas}, as there has been considerable activity in this
area in recent times.

\section{Instantons of Ho\v{r}ava--Lifshitz gravity}
\setcounter{equation}{0}

In this section we concentrate for definiteness to geometric flows as
Euclidean solutions of Ho\v{r}ava--Lifshitz gravity theories and provide explicit
examples of instanton configurations. We will consider models in $3+1$
space--time dimensions and choose the action of three--dimensional
topologically massive gravity as our superpotential functional, \cite{DJT},
\be
W = {2 \over \kappa_{\rm w}^2} \int d^3x \sqrt{g} ~ (R - 2 \Lambda) +
{1 \over \omega} W_{\rm CS} ~,
\ee
where
\be
W_{\rm CS} = \int d^3 x \, \sqrt{g} ~ \varepsilon^{ijk}
\Gamma{^\ell}_{im} \left( \partial_j \Gamma{^m}_{\ell k}
+\frac{2}{3} \Gamma{^m}_{jn} \Gamma{^n}_{k \ell}\right)
\ee
is written in terms of the usual Levi--Civita connection of the three--dimensional
metric $g$ and $\varepsilon^{ijk}$ is the fully anti-symmetric symbol
with $\varepsilon^{123}=1$. This choice yields a non--relativistic theory
of gravity in $3+1$ dimensions with anisotropic scaling $z=3$. As will be seen
shortly, going beyond $z=2$ is necessary in order to be able to produce
examples of instanton configurations in Ho\v{r}ava--Lifshitz gravity.

Let us briefly describe some of the salient features of topologically massive
gravity that will be needed in the following. The first term of $W$ is the usual
Einstein--Hilbert term, which is also augmented with a three--dimensional cosmological
constant $\Lambda$, whereas the second term is the so called gravitational
Chern--Simons action. The latter flips sign under orientation reversing transformations
and, as a result, the theory of topologically massive gravity is not invariant under
parity. The classical equations of motion that follow from $W$ by varying the metric
read as
\be
R_{ij} - \frac{1}{2} R g_{ij}
+ \Lambda g_{ij} + {\kappa_{\rm w}^2 \over \omega} C_{ij} = 0 \, ,
\label{soltmg}
\ee
where $C_{ij}$ is the Cotton tensor of the metric $g$, which is defined as follows
\be
C_{ij} = {{\varepsilon_i}^{k \ell} \over \sqrt{g}}
\nabla_k (R_{j \ell} - \frac{1}{4} R g_{j \ell})
\ee
and it is a traceless and covariantly conserved symmetric tensor. Thus, the
classical equations of motion contain second and third order derivative terms,
in general, explaining the scaling exponent $z=3$ in the associated
$(3+1)$--dimensional Ho\v{r}ava--Lifshitz gravity.

According to the general theory, Euclidean solutions of the Lifshitz theory on
$\mathbb{R} \times \Sigma_3$ are provided by trajectories of the corresponding
gradient flow equation which takes the form, \cite{bakas},
\begin{equation}
\partial_t g_{ij} = \mp \frac{\kappa^2}{\kappa_{\rm w}^2}\left(
R_{ij} - {2\lambda - 1 \over 2(3\lambda -1)} R g_{ij} - \frac{\Lambda}{3\lambda -1}
g_{ij} \right) \mp \frac{\kappa^2 }{\omega} C_{ij}
\end{equation}
keeping both sign options for completeness.
This is called Ricci--Cotton flow for the metric $g$ on $\Sigma_3$. Its fixed
points are independent of the parameter $\lambda$ and coincide with the classical
solutions of topologically massive gravity given by \eqn{soltmg}. Instanton
solutions exist provided that there are more than one fixed points and they
correspond to trajectories that interpolate smoothly between them as $t$ varies
from $-\infty$ to $+\infty$. It is certainly very difficult to derive general
mathematical results about the properties of the Ricci--Cotton flow because it is
a third order equation and the standard tools based on maximum principle do not
apply. Nevertheless, it is possible to make appropriate ansatz that reduce the
flow equation into a simpler system of ordinary differential equations, which
in turn can be studied in detail and yield explicit configurations.

We are led to consider locally homogeneous geometries on $\Sigma_3$, which for
simplicity it is assumed to have the topology of $S^3$, and make the Bianchi IX
ansatz for the three--dimensional metrics
\begin{equation}
ds^2 = \gamma_1 (t) \sigma_1^2 + \gamma_2 (t) \sigma_2^2 + \gamma_3 (t) \sigma_3^2
\end{equation}
so that the isometry group is $SU(2)$. Here, $\sigma_i$ are the left--invariant
one--forms of $SU(2)$ satisfying the defining relations
$2d\sigma_i + {\varepsilon_i}^{jk} \sigma_j \wedge \sigma_k = 0$. Remarkably,
this class of metrics provides consistent truncation of the Ricci--Cotton
flow, as for all other homogeneous model three--geometries that arise in Bianchi
classification. Sticking to this mini--superspace model, we will be able to
describe all instanton solutions of $z=3$ Ho\v{r}ava--Lifshitz gravity with
$SU(2)$ isometry, following \cite{bakas}. Since the instantons of Lifshitz
theories correspond to eternal solutions of the gradient flow equations, their
space--time interpretation is straightforward as they give rise to complete and
regular metrics
on $\mathbb{R} \times S^3$. Here, due to space limitation, we only present in
words the construction of these instantons and refer the reader to our
published work for the details. However, we are permitted to give one simple
example later to illustrate our general results on $SU(2)$ gravitational
instantons.

First, we need to characterize all fixed points, i.e., homogeneous vacuum
solutions of three--dimensional topologically massive gravity,
\cite{vuorio, nutku, ortiz} (but see also \cite{chow} for an overview).
It turns out that apart from the maximally
symmetric vacuum with $\gamma_1 = \gamma_2 = \gamma_3$, which always exists
for all values of parameters, there can be additional fixed points that are
formed by balancing the effect of the Einstein and Cotton terms to geometry.
Clearly the sign of the Chern--Simons coupling $\omega$ plays important role in
this problem as it distinguishes the case that the Einstein and Cotton
tensors compete against each other from the case that they work together at
the fixed points. Additional fixed points are present only in the first
case and they correspond to axially symmetric metrics on $S^3$. The exact
shape of the squashed or elongated spheres that can arise depends on the
actual values of the couplings. When $\Lambda = 0$, it is also possible to
have totally anisotropic fixed points beyond a critical value of $\omega$.
Then, the instanton solutions are configurations interpolating smoothly between
these fixed points and describe eternal solutions of the Ricci--Cotton
flow. Depending on the values of parameters, there are axially symmetric
deformations of the metric that connect the fixed points, in which case
the instantons have enhanced $SU(2) \times U(1)$ group of isometries, or
there can be more general instantons with $SU(2)$ isometry alone, which are,
however, difficult to describe in closed form. In any case, it is possible
to classify all such instanton solutions, compute their action and find
their moduli, as it is explained in the original work \cite{bakas}
when $\lambda < 1/3$ and $\Lambda \geq 0$.

An important property of these instantons is their chirality. Since the
Cotton tensor is odd under parity, orientation reversing transformations
on $S^3$ flip the sign of the coupling constant $\omega$. This has dramatic
effect on the formation of fixed points since the Einstein and Cotton
tensor will work together and not against each other. Thus, the instantons
of $z=3$ Ho\v{r}ava--Lifshitz gravity will cease to exist as they rely on
the existence of more than one fixed point to support themselves, which is
possible only for one sign of the coupling $\omega$ (assuming that
$\kappa_{\rm w}$ is finite). Chiral instantons are not common to field
theories, but here we have a model provided by Lifshitz theories that make
them possible. Another important point is the absence of instantons in
$z=2$ Ho\v{r}ava--Lifshitz gravity, which explains the need to
resort to models with higher anisotropy scaling exponent. In that case,
the gradient flow equation is the Ricci flow, which admits only one fixed
point on $S^3$, \cite{yau}, as consequence of the Poincar\'e conjecture. Thus,
there can be no gravitational instantons of the kind we are considering here.
Finally, one may also consider instantons of Ho\v{r}ava--Lifshitz gravity
with higher scaling exponent, say $z=4$, by choosing $W$ to be the action
of three--dimensional new massive gravity (and generalizations
thereof), \cite{BHT}. Homogeneous vacua with $SU(2)$ symmetry have been
systematically classified in this case, \cite{sourdis}, and, therefore,
gravitational instantons with $SU(2)$ isometry can be made available by
considering eternal solutions of a certain fourth order
geometric flow equation for Bianchi IX metrics. The details are lying
beyond the scope of this presentation.

In the remaining part of this section, we illustrate the results
given in \cite{bakas}, by presenting the simplest possible example
of an instanton solution of $z=3$ Ho\v{r}ava--Lifshitz gravity. We choose
$W$ to be the gravitational Chern--Simons action by dropping
the Einstein--Hilbert term in the limit $\kappa_{\rm w} \rightarrow \infty$.
Then, the underlying three--dimensional theory is the conformal rather than
the topologically massive gravity, \cite{horne}. The fixed points
obey $C_{ij} = 0$ and correspond to conformally flat metrics on $\Sigma_3$.
The associated Lifshitz theory is pure Cotton and its Euclidean solutions
are simply described by trajectories of the Cotton flow, which was also
introduced in \cite{tekin},
\begin{equation}
\partial_t g_{ij} = \mp \frac{\kappa^2 }{\omega} C_{ij} ~,
\end{equation}
and it is independent of $\lambda$. Since the Cotton tensor is traceless,
this deformation of the metric preserves the volume of space and the
calculations simplify a lot.

Within the mini--superspace model of Bianchi IX metrics on $S^3$, we are led
to consider axially symmetric configurations with fixed volume $2 \pi^2 L^3$
setting, in particular,
\begin{equation}
\gamma_1 = \gamma_2 \equiv x ~ {L^2 \over 4} ~, ~~~~~~
\gamma_3 = {L^2 \over 4x^2} ~.
\end{equation}
Then, the Cotton flow equation reduces consistently to a single equation for
the variable $x(t)$ that parametrizes the shape of space (squashing) and reads as
\begin{equation}
{dx \over dt} = \pm {4 \kappa^2 \over \omega L^3} ~ {x^3 - 1 \over x^5} ~.
\label{ramis}
\end{equation}
Clearly, there are two fixed points, one with $x=1$ and another with $x=\infty$.
They both describe conformally flat metrics on $S^3$ as $dx/dt$ vanishes there.
The first one corresponds to the round metric on $S^3$ having $\gamma_1 = \gamma_2
= \gamma_3$, whereas the second one arises in the correlated limit $\gamma_1 =
\gamma_2 = \infty$ and $\gamma_3 = 0$ with the volume held fixed. As such it looks
degenerate, since $S^3$ is completely flattened out, but it is non--singular (it
can be explicitly verified that it has no curvature singularities). These
are the two bona fide fixed points that can and will support an instanton.

Simple integration of equation \eqn{ramis} yields a branch with $x \geq 1$
that exists for all time and it is explicitly given by
\begin{equation}
\pm {t \over \tau} = x^3 - 1 + {\rm log} (x^3 -1) ~,
\label{cotain}
\end{equation}
where $\tau = |\omega| L^3/12\kappa^2$ is the characteristic time scale of the problem.
As $t$ ranges from $-\infty$ to $+\infty$, $x(t)$ interpolates smoothly between
the values $1$ and $\infty$ that describe the location of the two fixed points.
As noted before, the choice of sign distinguishes instantons from anti--instantons.
The instanton we have obtained in this fashion has finite Euclidean action, which
turns out to be $S_{\rm inst.} = 4\pi^2 / |\omega|$, and it has no modulus other
than $L$. It can also be seen without much effort that the solution \eqn{cotain} is the
only instanton with $SU(2)$ isometry (up to permutations of the three principal
axes of $S^3$) of the pure Cotton theory. In this case, the instanton and the
anti--instanton are simply interrelated by parity transformations on $S^3$.

The figure below serves to illustrate the situation. It is a plot of the potential
term $V = C_{ij} C^{ij}$ of pure Cotton $z=3$ Ho\v{r}ava--Lifshitz gravity for
axially symmetric Bianchi IX metrics on $S^3$, which turns out to be  proportional to
$(e^{-6 \beta_+} - e^{-3 \beta_+})^2$. The horizontal axis is the shape modulus
of the sphere, which is conveniently described here by the variable
$\beta_+ = {\rm log} x$; the notation is borrowed from mixmaster dynamics where
this parametrization is widely used, \cite{misner}, \cite{mixa}.
Thus, the two fixed points are located
at the origin and at infinity and clearly they are degenerate since the potential
vanishes there. There is also a small bump in between the two fixed points which is
peaked at $\beta_+ = ({\rm log} 2)/3$, where $\gamma_1 = \gamma_2 = 2 \gamma_3$.

\begin{figure}[h]\centering
\epsfxsize=13cm\epsfbox{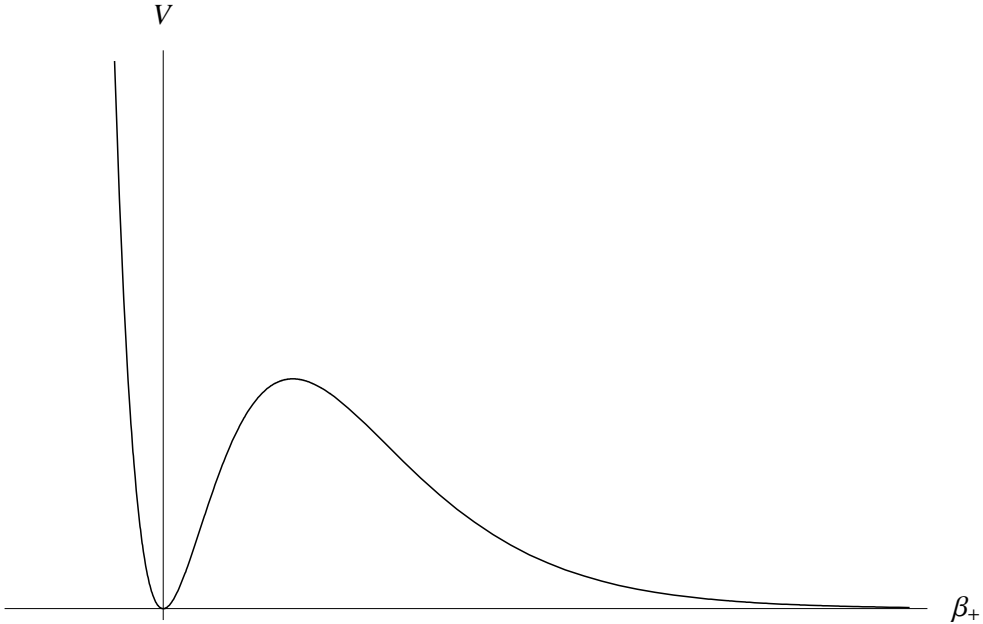}
\caption{Plot of the potential $C_{ij}C^{ij}$ for metrics with $SU(2) \times U(1)$
isometry.}
\end{figure}

The instanton interpolates between the two degenerate vacua, as in ordinary
particle systems. This analogy is made exact by noting that the Euclidean theory
of Ho\v{r}ava--Lifshitz gravity reduces in this case to the dynamics of a single
mode $\beta_+ (t)$ derived from the action
\begin{equation}
S_{\rm eff.} = {3 \pi^2 L^3 \over \kappa^2} \int dt \Big[\left({d \beta_+ \over dt}
\right)^2 + {16 \kappa^4 \over \omega^2 L^6} \left(e^{-6 \beta_+} - e^{-3 \beta_+}
\right)^2 \Big] ~.
\label{nairi}
\end{equation}
The change of variable $\beta_+ = {\rm log} x$ is necessary to cast the kinetic term
in canonical form. The instanton solutions of this effective point particle model
satisfy equation \eqn{ramis} written in terms of the variable $\beta_+$. Also, for
these configurations, the effective action equals $4\pi^2 / |\omega|$, as required for
consistency of the interpretation. The calculation is easily done by noting that the
potential term in \eqn{nairi} is itself derived from a superpotential, as
\begin{equation}
\left(e^{-6 \beta_+} - e^{-3 \beta_+} \right)^2 = \left({d W \over d \beta_+}
\right)^2 ; ~~~~~~~ W (\beta_+) = {1 \over 3} e^{-3 \beta_+} - {1 \over 6} e^{-6 \beta_+}
\end{equation}
so that
\begin{equation}
S_{\rm eff.}^{\rm inst.} = 2 ~ {3 \pi^2 L^3 \over \kappa^2} ~
{4 \kappa^2 \over |\omega| L^3}
~ |W (+\infty) - W(0)| = {4\pi^2 \over |\omega|} ~.
\end{equation}

Similar considerations apply to all other instanton solutions arising from consistent
truncation of the Ricci--Cotton flow to the Bianchi IX mini--superspace sector.
Such effective point particle systems, which are also commonly used in mixmaster dynamics,
prove useful for comparing $SU(2)$
gravitational instantons of Ho\v{r}ava--Lifshitz gravity with those of general relativity.
Recall that in $(3+1)$--dimensional Einstein gravity, the instantons are
defined as regular geometries with self--dual Riemann (or more generally Weyl) curvature
tensor. This property is not shared by the instantons of Ho\v{r}ava--Lifshitz gravity.
Yet the $SU(2)$ instantons of Einstein gravity admit an alternative description as
trajectories of a point particle moving under the influence of an effective potential,
\cite{pope}. The difference is that the metric in the space of truncated degrees of
freedom is indefinite (inherited by
the choice $\lambda =1$ in the Einsteinian DeWitt superspace metric) and the effective
potential does not exhibit degenerate vacua. In that case, the $SU(2)$ instantons
(such as the self--dual Taub--NUT solution) correspond to special trajectories of the
particle extending from a local maximum to a local minimum of the effective
potential. In terms of the four--dimensional geometry, they are supported by
removable (nut) singularities at one end, leading to
complete and everywhere regular space--time metrics. On the other hand, the
singularities that the effective point particle may encounter in Ho\v{r}ava--Lifshitz
gravity are not removable, which explains the need to have interpolating
trajectories between degenerate vacua to account for instantons in non--relativistic
gravitational theories. An important consequence of this difference is in the
asymptotic structure of the corresponding space--time metrics: asymptotically locally
flat metrics are only possible in Einstein gravity.

\section{Concluding remarks}
\setcounter{equation}{0}

The framework we have presented here is quite broad and can be used to provide a
toy model for dynamical vacuum selection in relativistic field theories
associated to an action $W$ in $d$ Euclidean dimensions. The instanton solutions
of the associated Lifshitz theory in $d+1$ space--time dimensions describe
off-shell transitions among the many different vacua that populate the landscape
of the $d$--dimensional theory. This alternative interpretation of Lifshitz models
is in the spirit of the Onsager--Machlup theory for non-equilibrium processes in
thermodynamics, \cite{onsa}, which is also based on the action \eqn{action} (or better
to say $S_{\rm Eucl.}^{\prime}$) and it is
often called Onsager--Machlup functional in the literature.
In this general context, $W$ is the entropy function that changes monotonically in time
and it is proportional to the logarithm of the probability of a given fluctuation.
The gradient of $W$ is the thermodynamic force measuring the tendency of a system to seek
equilibrium. Linearization of the flow equations around the fixed points describe
small fluctuations away from equilibrium states, whereas the instanton solutions
incorporate non-linear effects for large transitions between different equilibrium
states of the system. It will be interesting to strengthen the analogies between
non-equilibrium processes and gradient flow equations in the future.

Some aspects of our work are also reminiscent of renormalization
group equations, in particular for geometric theories, and we would like to understand
them better. Recall that transitions among vacua of string theory are often described
by running solutions of the beta--function equations. These off--shell processes are
themselves gradient flows derived from an effective gravitational action coupled to a
dilaton and possibly other massless fields of string theory. Thus, they can be
alternatively viewed as Euclidean solutions of a Lifshitz theory in one dimension higher
by identifying the renormalization group time with the extra Euclidean time dimension.
In this context, the off--shell dilaton field is treated as a Lifshitz scalar field, and,
likewise, all other off-shell massless fields are treated as Lifshitz tensor fields. However,
there is an important technical difference that prevent us from taking immediate advantage
of the results we described above. Here, we have been working under the assumption that
the metric in field space is positive definite (the same also applies to Onsager--Machlup
theory) in order to obtain instantons as extrema of the Euclidean action and interpret
$W$ as entropy functional. On the other hand, the renormalization group equations
used in the off--shell formulation of string theory follow from an indefinite metric by
choosing, in particular, $\lambda = 1/2$ in DeWitt's superspace metric (it is the same
choice that turns \eqn{riccia} into the ordinary Ricci flow equation for the pure metric
sector of the theory). Of course, one
can adjust the value of $\lambda$ accordingly to make this metric positive definite, thus
taking advantage of all goodies that come with it, and then
use the resulting equations as toy model for dynamical vacuum selection in string
theory. This modification does not affect the structure of the fixed points, whose
defining relations are inert to $\lambda$, but the flow lines will not be trajectories
of the renormalization group equations. It remains to be seen what else can be learned
from this analogy.

Another aspect of the present work that is potentially interesting for mathematics
is the unifying framework that Lifshitz theories provide to all gradient flow equations.
Our results on Ho\v{r}ava--Lifshitz theories in $3+1$ dimensions also provide a good
reason to study higher order geometric flows more systematically. The Cotton flow is
the simplest example of this kind, being an equation of third order, but one should also
study more systematically equations of mixed order such as the Ricci--Cotton flow. These
systems possess an abundance of fixed points, and, hence, they have much richer structure
than the ordinary second order Ricci flow allowing for eternal (instanton) solutions.
The main technical problem is to develop methods to estimate curvature bounds and obtain
general criteria for the formation of singularities along such flow lines. These issues
remain largely unexplored at this time and they call for immediate attention by the experts.

Finally, we comment on the possible use of our instanton solutions to Ho\v{r}ava--Lifshitz
theory when viewed as an alternative theory of gravitation at short distances. It
will be interesting to obtain a Euclidean path integral formulation of such a
non--relativistic theory of gravitation, where our instanton configurations will contribute
substantially, since they are weighted by the exponential of (minus) their action. Also,
mini--superspace models of quantum cosmology will be interesting to investigate
in this context by focusing, in particular, to the mixmaster universe based on the
homogeneous three--geometries, \cite{mixa}. Comparison with ordinary gravity may
provide some valuable lessons. Of course, it still remains to settle some open
questions regarding the general validity of Ho\v{r}ava's theory, which appears to
contain an unphysical scalar graviton mode, and its consistency with large scale
gravitational physics. Also, the condition of detailed balance on which our instanton
constructions
were based appears to be rather restrictive on physical grounds, as it does not seem to
allow for asymptotically flat solutions. A generalization of the original theory
was recently made to circumvent some of these problems, \cite{horava3}, and it might
still take time to settle the situation one way or another.

\vskip1.5cm
\noindent
{\Large{\bf Acknowledgements}}

\noindent
I am very grateful to the conference organizers for their kind invitation and
financial support. I also thank Francois Bourliot, Dieter L\"ust, Marios
Petropoulos as well as Christos Sourdis for fruitful collaboration on the main
aspects of the present work.

\end{document}